\documentclass{aa}
\usepackage{graphicx}
\usepackage{natbib}

\usepackage{txfonts}

\begin{document}
   \title{Formation and destruction of polycyclic aromatic hydrocarbon 
clusters in the interstellar medium}
 
   \author{M. Rapacioli,
          \inst{1}
F. Calvo,
\inst{2}
C. Joblin,
\inst{1}
P. Parneix,
\inst{3}
D. Toublanc
\inst{1}
          \and
F. Spiegelman
          \inst{2} 
          }

   \offprints{C. Joblin}

   \institute{Centre d'\'Etude Spatiale des Rayonnements,
     CNRS-Universit\'e Paul Sabatier, Observatoire Midi-Pyr\'en\'ees, 9
     Av. du colonel Roche, BP 4346, 31028 Toulouse Cedex 4, France\\
\and
     Laboratoire de Chimie et Physique Quantiques, IRSAMC, Universit\'e
     Paul Sabatier, 118 Route de Narbonne, 31062 Toulouse Cedex, France\\
\and
     Laboratoire de Photophysique Mol\'{e}culaire,  CNRS B\^{a}t 210, 
     Universit\'{e} Paris-Sud, 91405 Orsay Cedex, France
             }

   \date{Received April 12, 2006; accepted June 22, 2006}

 \abstract
{}
{The competition between the formation and destruction of coronene
  clusters under interstellar conditions is investigated
  theoretically.}
{The unimolecular nucleation of  neutral clusters is simulated
  with an atomic model combining an explicit classical force field
  and a quantum tight-binding approach. Evaporation rates are
  calculated in the framework of the phase space theory and are inserted
  in an infrared emission model and compared with the growth rate
  constants.}
{It is found that, in interstellar conditions, most collisions lead
  to cluster growth. The time evolution of small clusters (containing
  up to 312 carbon atoms) was specifically investigated under the
  physical conditions of the northern photodissociation region of
  NGC~7023. These clusters are found to be thermally photoevaporated 
  much
  faster than they are reformed, thus providing an interpretation
  for the lowest limit of the interstellar cluster size distribution
  inferred from observations. The effects of ionizing the clusters
  and density heterogeneities are also considered. Based on our
  results, the possibility that PAH clusters could be formed
  in PDRs is critically discussed.}
   
   \authorrunning{M. Rapacioli et al.}
   \titlerunning{Formation and destruction of PAH clusters}

   \maketitle
%

\section{Introduction}

Many astronomical objects show a distinct set of emission bands
in the mid-infrared range, known collectively as the  unidentified
infrared emission features. The most intense of these features,
the Aromatic Infrared Bands (AIBs) falling at 3.3, 6.2, ``7.7'',
8.6, 11.3 and 12.7~$\mu$m, are the signatures of aromatic CC
and CH bonds. They are observed systematically from different regions
of the interstellar medium (ISM) irradiated by UV photons. The carriers
of these bands have been identified as Polycyclic Aromatic
Hydrocarbons (PAHs) some twenty years ago by \cite{leger84} and
\cite{allamandola85}. 

\cite{Boulanger90} and \cite{Bernard93} analysed the emission
in the IRAS photometric bands measured for several molecular clouds,
and suggested that free-flying PAHs are produced by photoevaporation
of larger grains. The observation
of a mid-infrared continuum in the reflection nebula Ced 201 has been
attributed by \cite{Cesarsky00} to Very Small carbonaceous Grains
(VSGs) producing the AIB carriers. More recently, in their study of
the two reflection nebulae NGC 7023 and $\rho$ Ophiucus-SR3,  
\cite{rapacioli05a} showed that PAH molecules are produced by the destruction
of small carbonaceous grains inside molecular clouds, these grains
being interpreted as PAH clusters. A minimal size of 400 carbon atoms per
cluster was inferred from the analysis of astronomical observations
\citep{rapacioli05a}. A plausible scenario is that these grains undergo
photo-evaporation at the edge of clouds, thus leading to isolated PAHs. 
These observations motivate theoretical and experimental
studies on the physical properties of PAH clusters, such as
structure, stability or aggregation.

Small clusters of PAH molecules have been studied by several authors
in the past, both experimentally
\citep{benharash,song,piuzzi,miller84}
and theoretically
\citep{vandewaal1,gonzalez99,marzec,piuzzi,grimme}. Most of these studies have
focused on rather small species,  usually containing no more than a
few tens of carbon atoms, far below the limit of 400 carbon atoms
per VSG as suggested by \cite{rapacioli05a}. 
Studies on larger systems have been initiated only lately,
motivated by the astrophysical context. 
Clusters of coronene molecules containing up to 20 units 
have been produced in a gas aggregation source 
\citep{brechignac05, schmidt06}. Interestingly, the authors 
showed that the dissociation of coronene clusters induced by 
multi-photon excitation of a near-UV laser proceeds via the ejection 
of van der Waals-bonded coronene units. 

In a previous work \citep{rapacioli05b}, we investigated the stable
structures of PAH clusters containing up to 800 carbon atoms (32
coronene molecules). The
considered PAHs ranged from pyrene (C$_{16}$H$_{10}$) to
circumcoronene (C$_{54}$H$_{18}$). The most stable cluster structures
generally result from stacking the PAH molecules, first in a one-dimensional
pattern at small cluster sizes. Above some critical size (depending on
the PAH itself), structures more stable than the single stack are
found, consisting of shorter stacks lying next to each other, and
growing as two-, eventually three-dimensional structures. In large
PAH clusters, the multiple ways of arranging the short stacks lead to
a competition between stable structures.

Our present interest lies in the formation and destruction
of PAH clusters in the ISM, and particularly the competition between these
two processes. The coronene molecule (C$_{24}$H$_{12}$) provides a
convenient example of large PAHs, especially from the computational
point of view, and is used  in the present work as a prototype for
interstellar PAHs. We wish to address here the following questions:
\begin{itemize}
\item[(i)]  What are the astrophysical conditions that favour
  the nucleation of PAH clusters?
\item[(ii)] How much energy is required for thermal evaporation of these
  clusters?
\item[(iii)] How do the rates of formation and destruction
compare for these clusters under interstellar conditions?
\end{itemize}

In this paper, we focus on small neutral coronene clusters
containing up to 312 carbon atoms, or 13 molecules. In particular, we
attempt to offer a physical explanation for the
existence of the observed lower limit to the interstellar PAH cluster
size distribution of 400 carbon atoms per cluster \citep{rapacioli05a}.


Addressing these questions requires a variety of
theoretical methods to be used. Nucleation processes are explicitly
simulated with an extended version of the tight-binding (TB) model 
initially developed by \cite{parneix} for single PAHs.
Concerning the thermal stability of coronene
clusters, the rates of unimolecular evaporation as well as the
photoemission rates in the infrared are obtained and compared. 
The evaporation rates are obtained in the statistical
framework of the phase space theory \citep{bowers77}. The competition
between evaporation and IR emission is calculated using a master
equation modeling based on microcanonical statistics \citep{joblin02}.
In the light of these calculations, we discuss several chemical
scenarios to interpret the observations of the reflection nebulae
\citep{rapacioli05a}.

The paper is organized as follows. In the next section, we describe
the atomic model and discuss the numerical simulation of coronene
clusters growth by unimolecular nucleation. In section
\ref{sec:relax}, the evaporation rates of coronene clusters are
calculated. In section \ref{sec:compet}, we study the competition
between evaporation and IR emission. In section \ref{sec:photophys},
the photophysics of a coronene cluster in a radiation field is
simulated. In section \ref{sec:balance}, we study the competition
between formation and destruction of coronene clusters under
interstellar conditions. We also highlight the astrophysical
implications of our results for possible models of
chemical evolution in the photodissociation regions. In
particular, the possible roles of ionization and density
heterogeneities in the PDR in the existence and stability of PAH
clusters are discussed. Some conluding remarks are given in
Sec.~\ref{sec:con}.


\section{Cluster formation process}
\label{sec:formation}

In this section, we consider the formation of  neutral coronene
clusters by collision of a single molecule with a smaller cluster or
with another single molecule. In the ISM, collisions occur in a very
rarefied gas and the time lapse between successive UV absorption events is
in general much longer than the IR emission timescale. Therefore the
PAH molecules and their clusters have enough time to cool down through
IR emission, and can be considered as initially cold in their ground
electronic and vibrational states. 
 This energy will be transferred to intermolecular modes,
then to the intramolecular modes. Intramolecular vibrational energy
redistribution (IVR) is likely to be the major source for dissipating
the initial energy, having thus strong consequences on the
collision products. Hence it is important to take this redistribution into account in
molecular simulations.

In this paper, we will consider that a new cluster is formed upon collision
if it has time to reach thermal equilibrium before further
dissociating. The relaxation of a thermalised cluster heated by
collision or by photoabsorption is
discussed in section \ref{sec:relax}. We describe hereafter the
atomic model that was implemented for the present problem.

\subsection{Atomic model}
\label{sec:atmod}

Describing the transfer of the collisional energy toward
intermolecular and intramolecular modes requires both types of degrees of freedom to
be correctly accounted for. Clusters of coronene molecules are modeled
using the combination of an explicit classical force field to describe the
intermolecular interactions, along with a quantum tight-binding model for the
intramolecular interactions within each molecule. The total
Hamiltonian for a system with $N_{\rm mol}$ PAH molecules is written:
\begin{equation}
  H=\sum_{i=1}^{N_{\rm mol}} H^i_{\rm intra}+H_{\rm inter}.
\end{equation}
In this equation $H^i_{\rm intra}$ is the intramolecular Hamiltonian of
molecule $i$, which only explicitly depends on the positions
of the atoms in this molecule. We use for $H^i_{\rm intra}$ the
tight-binding model developed by \cite{parneix}.
This simple quantum model was parameterized to reproduce energetic and
vibrational properties of isolated PAH molecules.
We did not change its original parameters.

$H_{\rm inter}$ is the intermolecular potential resulting from
the repulsion-dispersion forces as well as the electrostatic interactions
between partial charges on the atoms of different molecules. 
Following our previous work \citep{rapacioli05b}, we use a simple
Lennard-Jones (LJ) + point charges potential, the charges being determined
to reproduce Density Functional Theory (DFT) 
calculations on an isolated molecule. The LJ parameters are
taken from \cite{vandewaal1}. Further details about the intermolecular
potential, including a discussion about the sensitivity of the results
with respect to the potential parameters, can be found in \cite{rapacioli05b}.

\subsection{Simulation details}

The classical equations of motion on the Born-Oppenheimer potential
energy surface have been numerically integrated using the fifth-order
Adams-Moulton predictor-corrector algorithm. The time step was chosen
between 0.05 fs and 0.15 fs depending on the collision energy. 
The computational effort required to diagonalize the multiple TB Hamiltonians 
was distributed among many processors on parallel computers in the OpenMP framework.
As stated previously, the reactants are assumed to be cold, 
their vibrational energies being taken as less than 80~K (much lower than the
collisional energy).
The physical parameters, which characterize a collision in the centre of
mass reference frame, are the collisional energy (translational plus
rotational components), the impact parameter,
the angular momentum of each fragment and their initial orientations.
The statistical information obtained from the simulations was limited
due to the still relatively heavy numerical cost involved in computing the
TB energies and forces. Therefore we restricted our study to the
following representative situations, assuming this restriction has a
minor impact on the collision products:
\begin{itemize}
\item all collisions are at zero impact parameter (no orbital angular
  momentum);
\item the total angular momentum of the system (incident
  molecule+cluster) is zero;
\item the collisional energy is equally distributed between the
  translational energy and the rotational energy of the reactants.
\end{itemize}
At a given collisional energy, the free mechanical parameters are thus
reduced to the initial orientations of the molecule and the cluster
and the direction of the angular momentum of the cluster. We did not
attempt to determine quantitative cross sections for the collision
processes.

About 30 trajectories have been performed for each collisional energy 
in order to sample these parameters and to gather statistics on the
outcome of the collisions.  For simplicity the collision of reactants
 with sizes $p$ and $p'$, respectively, leading to a set of
fragments with sizes $d$, $d'$, $d''$... will be referred to as $p+p'\to
d+d'+d''+\cdots$, where the parents and products are
ordered by increasing size.

In practice the simulation ends when either one of the two following situations happens:
\begin{itemize}
\item after the collision, fragmentation occurs and one or more
  molecules are ejected far from the others;
\item the instantaneous intra- and intermolecular temperatures cross
  each other, meaning that the product cluster has absorbed the
  collisional energy and has reached thermal equilibrium.
\end{itemize}
Other situations have not been observed.

\subsection{Results}
\label{sec:resag}

\begin{figure}[htbp]
\begin{center}
\includegraphics[width=9 cm]{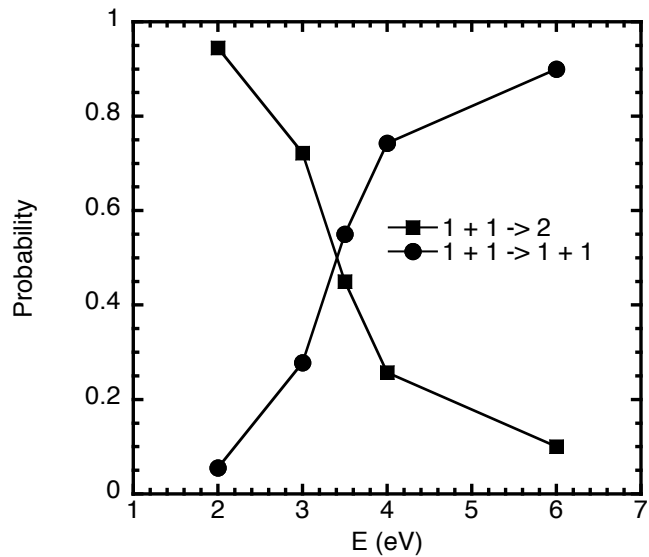}
\end{center}
\caption{Probabilities $p^0_+$ and $p^0_n$ that a collision between two 
coronene molecules forms a dimer ($1+1\to 2$) or not ($1+1\to 1+1$),
as a function of the collisional energy. Each data point is averaged over
30 simulations. Lines are drawn to guide the eye.}
\label{fig:resformdi}
\end{figure}

Figure \ref{fig:resformdi} shows the nucleation probability of
simulated collisions between two coronene molecules as a function of
collisional energy, that is the chance that a collision leads to the
formation of a dimer as $1+1\to 2$. The probability that the collision
fails in producing a dimer ($1+1\to 1+1$), also plotted in
Fig. \ref{fig:resformdi}, complements the nucleation
probability.

These results show that the probability of forming a dimer
decreases with increasing collisional energy. This is not surprising,
since collisions at higher energy tend to make the two molecules
bounce of each other. Nucleation is efficient,
especially at low energies. For instance, the probability of forming a
coronene dimer exceeds 50\% when the collisional energy is lower
than 3.6 eV. This rather high energy threshold corresponds to the
average collisional energy of a gas at a temperature of 13 500~K.

\begin{figure}[htbp]
\begin{center}
\includegraphics[width=9 cm]{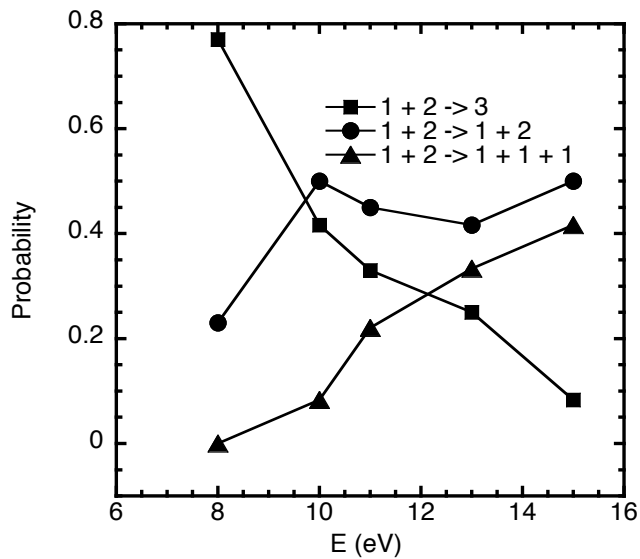}
\end{center}
\caption{Probabilities $p^0_+$, $p^0_n$, and $p^0_-$ that a collision
of a single coronene on a dimer forms a trimer ($1+2\to 3$), leaves a
dimer ($1+2\to 1+2$), or fragments into 3 molecules ($1+2\to 1+1+1$),
as a function of the collisional energy. Each data point is averaged over
30 simulations. Lines are drawn to guide the eye.}
\label{fig:resformtri}
\end{figure}

\begin{figure}[htbp]
\begin{center}
\includegraphics[width=9 cm]{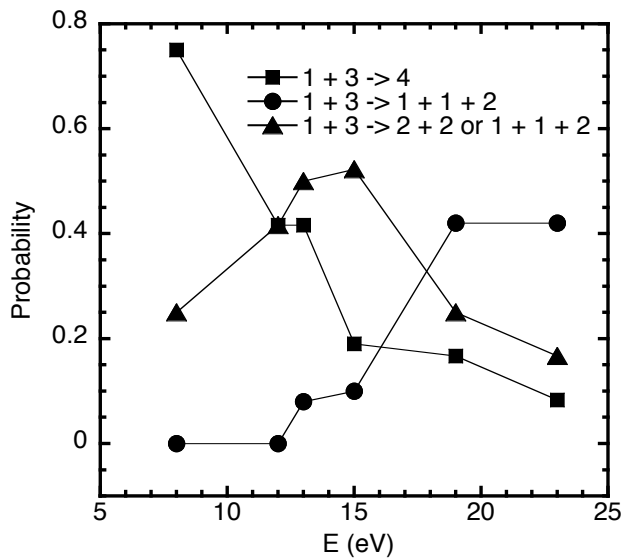}
\end{center}
\caption{Probabilities $p^0_+$, $p^0_n$, and $p^0_-$ that a collision
of a single coronene on a trimer forms a tetramer ($1+3\to 4$), leaves a
trimer ($1+3\to 1+3$), destroys the inital cluster ($1+3\to 2+2$ or
$1+3\to 1+1+2$), as a function of the collisional energy. Each data point
is averaged over 30 simulations. Lines are drawn to guide the eye. }
\label{fig:resformquat}
\end{figure}

The probabilities of successfully forming a larger cluster or of
destroying the initial cluster from a collision between a single
coronene and a stack of 2 or 3 molecules are shown in 
Figs. \ref{fig:resformtri} and \ref{fig:resformquat}, respectively.
For these clusters, the probability of a neutral event, that is when the
collision products are identical to the parents in terms of their
size, is also plotted. The multifragmentation of the collision product
into single molecules only was not observed in the case of the tetramer.

As in the case of dimer formation, the probability of growing
the initial cluster decreases monotonically with increasing collisional
energy. A monotonic increase is also observed for the probability of
ending with smaller clusters. This allows us to define a critical
energy for which these two probabilities are the same.
This energy approximately equals 12.4~eV for the $1+2$
reaction and 14.2~eV for $1+3$. The increase of the critical energy
with cluster size is expected since a larger number of modes are
available to convert the collisional energy. 
 In our simulations, the range of collisional energies was not
chosen on the basis of astrophysical values but only to allow enough
statistics to be gathered in the calculations. Even though the mean
collisional energies encoutered in the interstellar medium are usually
much lower (typically in the 0.01--0.1~eV range), we extrapolate below
the values corresponding to the smallest calculated collisional energies
to collisions with lower energies. This approximation leads to some
underestimation of the cluster growth, which we estimate to be no more
than 25\% since the sticking probability is at least 0.75 (as seen
from Figs.~\ref{fig:resformdi}--\ref{fig:resformquat}).

\section{Evaporation of a neutral isolated cluster}
\label{sec:relax}

A thermalised isolated coronene cluster with an excitation energy
$E_{\rm tot}$ can release a part of this excess energy via
dissociation, evaporation and IR emission.
We will assume that the cluster remains on the ground state electronic
potential surface. In this section, we study the evaporation mechanism,
which in turn requires an estimation of the characteristic time for the energy
redistribution between inter- and intramolecular modes.
In the next section, we study the competition between the different
relaxation mechanisms involving molecular decay or infrared emission.

\subsection{Vibrational redistribution}

Neglecting the influence of the excited states, 
collisions and UV-visible photon absorption under astrophysical conditions 
leads to an increase of the internal energy of the cluster, which 
can then be dissipated either by evaporation and/or IR emission.
The energy redistribution between inter- and intramolecular
vibrational modes is required for the cluster to be considered
in thermal equilibrium upon nucleation of a molecule. 
Similarly, the absorption of a photon by a single molecule in the
cluster will tend to heat the intramolecular modes of this specific
molecule first by internal conversion, before the excess energy flows
into intermolecular modes, and subsequently into the intramolecular modes of
the other  molecules. The time scales involved in these fundamental
redistribution processes should be considered as reference data in the
next stages of our investigation.

As a first application of our model for PAH clusters, we have
simulated the energy redistribution of intramolecular energy toward
intermolecular modes by heating a single molecule in the
cluster and monitoring the amount of kinetic energy in all other
molecules as a function of time. The kinetic energy provides
a direct estimate of the vibrational or thermal energy.
We consider redistribution in the
(coronene)$_5$ cluster as a typical example. This cluster
consists of a short stack, for which either the outermost or the
central molecule are chosen to carry the initial thermal excitation.
The starting configuration of the cluster is taken as the lowest-energy
structure, and no extra kinetic energy is given to the intermolecular modes.
To quantify the redistribution of energy into vibrational modes of all
molecules in the cluster, we introduce the root mean square fluctuation
$\delta_T$ of the kinetic temperatures of the molecules in the cluster:
\begin{equation}
  \delta_T(t)=\frac{1}{N_{\rm mol}}\sqrt{\sum_{i=1}^{N_{\rm mol}}
  \left( T_i(t)-\langle T(t)\rangle \right)^2}
\end{equation}
where $T_i(t)$ is the average kinetic temperature restricted to
molecule $i$ and $\langle T(t)\rangle$ is the usual average kinetic
temperature of the entire system.  $\delta_T(t)$ provides an
instantaneous measure of the standard deviation of the individual
temperatures  of all molecules in the cluster. Initially
this quantity is large, since only one molecule carries kinetic
energy. Eventually the cluster reaches equilibrium and the
temperatures of all molecules in the cluster are identical. The
rate at which $\delta_T(t)$ decreases yields an estimate for the time
scale $\tau$ of energy redistribution
in the system, according to $\delta_T(t) \sim \delta_T(0)e^{-t/\tau}$.

\begin{figure}[htbp]
\begin{center}
\includegraphics[width=9 cm]{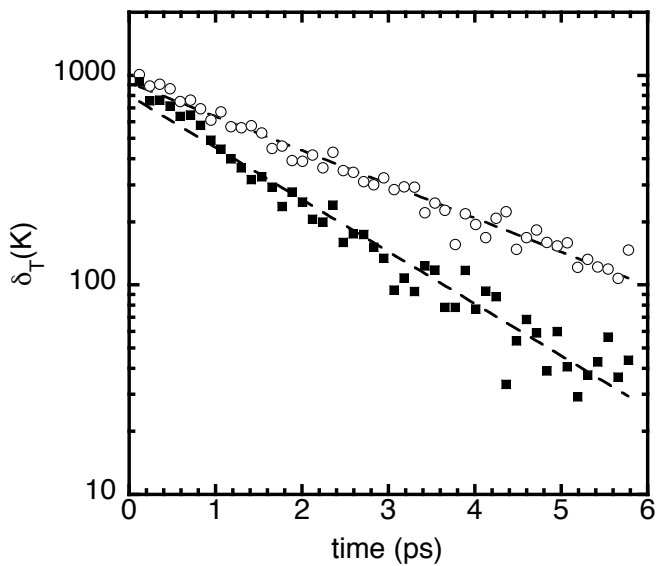}
\end{center}
\caption{ Root mean squares $\delta_T$ of the kinetic energies
 of the
  molecules in a   cluster of five molecules. At $t=0$, the initially
  kinetic energy excess (13 eV) is   localized in the outermost
  (circles) or central (squares) molecule   within the stack. 
The dashed lines are linear fits.}
\label{fig:vibdist}
\end{figure} 

Figure \ref{fig:vibdist} shows the variations of $\delta_T$ as a function
of time for (C$_{24}$H$_{12}$)$_5$ in which the initial
excitation is localized on different molecules. The decrease of
$\delta_T$ is indeed exponential, with decay time constants approximately 
equal to $\tau=1.47\pm 0.09$ ps when the central molecule is vibrationally
excited, and $\tau=2.67\pm 0.28$ ps
when the outermost molecule is excited.
These time scales are of the same magnitude as the softest vibrational
modes of isolated coronene molecules \citep{martin96b}. This is
not surprising, because the slowest modes constitute a lower limit
to the rate at which energy is redistributed homogeneously within all
internal degrees of freedom. Finally, the faster redistribution when the
central molecule is heated is expected due to the presence of fewer
intermolecular modes involved.

\subsection{Evaporation}

Coronene clusters could {\em a priori} dissociate into entire
molecules, or via intramolecular fragmentation.
The loss of a molecule from a coronene cluster
[typically 1~eV of binding energy \citep{rapacioli05b}] is much more
likely to happen than the breaking of chemical bonds, which requires
more than 4~eV \citep{joblin06}.
In agreement with the experimental 
conclusions of \cite{schmidt06}, we thus only consider
here the dissociation corresponding to single molecule evaporation.

Evaporation rates of molecular clusters are not available by direct
molecular dynamics simulations except at very high excitation
energy. When the extra energy deposited in the cluster is marginally
higher than the dissociation energy, fragmentation is a rare event,
which can take place over time scales much longer than the typical
times that are reachable by conventional molecular simulations.
Statistical rate theories are much more convenient for calculating such rates. 
The phase space theory (PST)  developed by \cite{bowers77} 
provides an accurate framework for unimolecular dissociation.
In previous work, two of us \citep{calvo03c} showed
that PST predictions were in quantitative agreement with numerical
simulations for molecular systems, provided that the rotational and
vibrational densities of states (DOS) were correctly evaluated.


\subsubsection{Phase space theory}

The success of PST is essentially due to the rigorous conservation 
of energy and angular momentum, and to the inclusion 
of anharmonic
vibrational effects. Here we are mainly interested in the
determination of absolute rate constants for the evaporation of a
single molecule from a thermalized cluster. PST expressions for
the rate constant \citep{jarroldrev} contain a set of factors
that are not always obvious to estimate with accuracy. Instead of
assuming specific values for these unknown factors, we follow
\cite{amar} and use the results of molecular dynamics simulations
at high internal energy to calibrate these unknown factors.

Performing molecular dynamics simulation of the cluster with both
inter- and intramolecular interactions is not realistic in the context
of thousands of trajectories over time scales governed by the
intermolecular motion. Therefore we had to simplify the modeling by
freezing intramolecular motion. With this assumption, the molecules
are treated as rigid bodies, and larger time steps and algorithmic
improvements such as the use of quaternion coordinates allow an
efficient simulation of the evaporation process. 
Even though the rigid body approximation may not be physically relevant 
at high energies, it calibrates the statistical theory, 
that we will subsequently use in a lower excitation regime.

In this subsection, energies are thus measured on the
intermolecular potential energy surface $H_{\rm inter}$ only. 
It is not our goal here to describe the PST formalism in
great details, and the reader is refered to the work by \cite{bowers77},
\cite{amar}, \cite{calvo03c} for further
information. The differential rate of unimolecular evaporation from a
molecular cluster $X_{n+1} \to$ X$_n+$X (here X=C$_{24}$H$_{12}$), at
total rovibrational energy $E$ and angular momentum $J$, and leading
to the loss of kinetic energy $\varepsilon_{\rm tr}$ into translation $\varepsilon_{\rm t}$
and
rotation $\varepsilon_{\rm r}$ is expressed as
\begin{equation}
  R(E,J,\varepsilon_{\rm tr})=C_0\frac{\Omega'(E-E_0-\varepsilon_{\rm tr})
    \Gamma'(J,\varepsilon_{\rm tr})}{\Omega(E-E_{\rm r})}.
\label{eq:difrate}
\end{equation} 
In the above equation $E_0$ is the dissociation energy, that is the
binding energy difference between the parent and the product
clusters, $E_{\rm r}$ is the parent rotational energy, equal to $BJ^2$ where
$B$ is the rotational constant. $\Omega$ and $\Omega'$ are the
vibrational densities of states of the parent and product clusters,
respectively. $\Gamma'$ is the rotational density of states of the two
products, obtained from the mechanical constraints on angular momentum at
a given translational+rotational energy. An estimation of $\Gamma'$ requires
treating the products as rigid bodies. $C_0$ is a constant
that we avoid calculating by performing high energy molecular dynamics
simulations for calibration.  The rate constant $k(E,J)$ is obtained
from the differential rate $R(E,J,\varepsilon_{\rm tr})$ by
integration over the available range of $\varepsilon_{\rm tr}$.

The ingredients of the PST calculations are threefold. First, the
vibrational densities of states $\Omega$ and $\Omega'$
are calculated using parallel
tempering Monte Carlo simulations \citep{neirotti}  analysed with
a multiple histogram method. These calculations consist of running
independent Monte Carlo trajectories or replicas at different
temperatures, occasionnally attempting to swap configurations between
adjacent replicas. Here we use 50 different temperatures in the range
10~K--1500~K distributed according to a geometric progression. For each
trajectory, $6~10^5$ Monte Carlo cycles (1 MC cycle=$n$
individual steps) are carried out for the calculation, the first $10^5$
steps being discarded for equilibration. 

The rotational density of states $\Gamma'$ requires integrating the
number of rotational states $\chi$ with specified internal angular
momentum of the products $J_r=J_1+J_2$ and specified orbital momentum
$L$ under the constraint that the translational energy released $\varepsilon_{\rm t}$ is larger that 
the centrifugal barrier $\varepsilon^\dagger$. The calculation becomes
significantly simpler by assuming that the initial angular momentum is
small or, more precisely, that the rotational energy is small with
respect to the vibrational energy. In that case $J_r=L$ in modulus,
and the integration is one-dimensional over the range of available
values of $J_r$:
\begin{equation}
\Gamma'(\varepsilon_{\rm tr},J\simeq 0)=\int_0^{J_r^{\rm
    max}}\chi'(\varepsilon_{\rm r}^*,J_r)dJ_r.
\label{eq:rdos}
\end{equation}
In the previous equation $\varepsilon_{\rm r}^*=\varepsilon_{\rm tr} -
\varepsilon^\dagger$ with $\varepsilon^\dagger(J_r)$ the energy of the
centrifugal barrier.

The function $\chi$ to be integrated has been calculated exactly by
\cite{bowers77} in a number of situations for products with
various shapes.
However, coronene is a highly oblate molecule, and expressions
are only available if the other product has a high symmetry. Therefore we
had to restrict our study to sizes for which the product
cluster is roughly spherical. From our previous results on cluster
structure \citep{rapacioli05b}, two sizes fulfill this criterion to a
good approximation, namely $n=3$ 
($B\approx 3.8~10^{-5}$~cm$^{-1}$) and $n=12$
($B\approx 1.4~10^{-6}$~cm$^{-1}$). 
Thus our investigation of unimolecular dissociation will focus on the
two parent clusters containing 4 or 13 molecules.

As a third ingredient, the PST calculation also needs an expression
for the dissociation potential between the products, in order to
locate the centrifugal barrier and estimate its height $\varepsilon^\dagger$.
At long range $r\to\infty$ the interaction between a cluster of PAH molecules and a
single PAH is essentially of the dispersion 
form, which we correct at short distances using the expression
 $-C_6/(r-r_0)^6$.
The values of $C_6$ and $r_0$ were obtained from a series of Monte Carlo
simulations with the distance between the two products constrained to
increasing values. We find $C_6=-4.3617~10^6$ kJ~\AA$^6$/mol 
(resp. $C_6=-1.6574~10^7$  kJ~\AA$^6$/mol) and 
$r_0=2.27$~\AA\ (resp. $r_0=3.73$~\AA) for $n=4$ (resp. $n=13$). 

\subsubsection{Calibration}

The statistical theory described previously can be used to calculate
kinetic energy release (KER) distributions, given by 
Eq.~(\ref{eq:difrate}) after normalization over $\varepsilon_{\rm tr}$. The KER has the advantage
over the rate constant that it does not include the unknown factor
$C_0$, hence it provides a good testing ground for the accuracy of
PST.

For each of the two parent clusters, we performed a sample of
10{\,}000 molecular dynamics simulations at relatively high excess
energy. For each trajectory, the cluster was initially prepared cold
(10~K), and was suddenly heated. The simulation was ended when
dissociation of a single molecule occured within 100~ps. Once
evaporation was detected, the dissociation time was taken as the
latest moment when the radial velocity of the dissociating molecule
was pointing inward. Here the MD results can be considered as numerically
exact, and as a reference for assessing the PST approach.

\begin{figure}[htbp]
\includegraphics[width=9 cm] {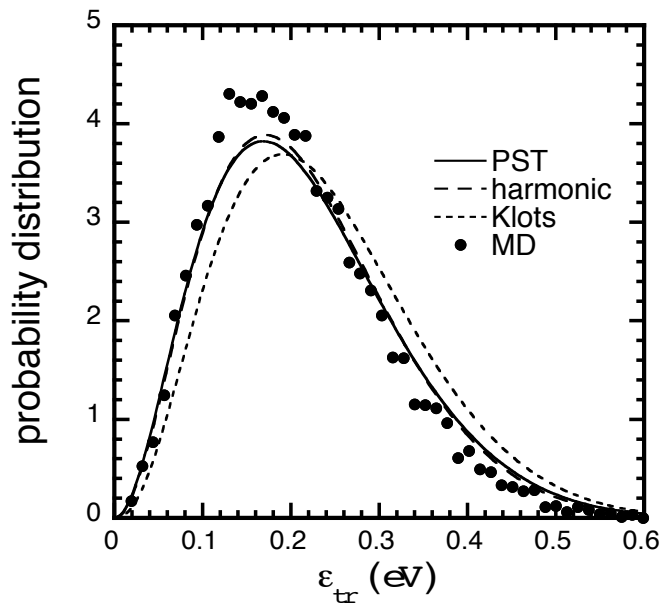}
\caption{Probability distribution of the kinetic energy released during 
the unimolecular evaporation of (C$_{24}$H$_{12}$)$_4$ 
at 2 eV 
intermolecular energy obtained 
from MD simulations and from PST based on several approximations (see the text for details 
about these approximations).
}
\label{fig:distribE}
\end{figure}

The distributions of KER obtained from the MD simulations are compared
with PST predictions in Fig.~\ref{fig:distribE} for the dissociation
of (C$_{24}$H$_{12}$)$_4$ at 2~eV excess energy. 

The PST calculations were performed at three different levels of
approximation. In its most rigorous form, the vibrational density of
states is the anharmonic form extracted from Monte Carlo simulations,
and the rotational density results from integrating Eq.~(\ref{eq:rdos})
above. The harmonic approximation for the vibrational density is
simply $\Omega(E)\propto E^{\nu-1}$ with $\nu=6n-6$ the number of
independent degrees of freedom in the system (3 rotational and 3 translational 
modes per molecule). The Klots approximation
at low angular momentum [see \citep{klots,calvo03c} for more details] 
is similarly $\Gamma'(\varepsilon_{\rm tr})\propto
\varepsilon_{\rm tr}^\alpha$ with $\alpha=(r-1)/2$, $r=6$ being the number
of rotational degrees of freedom in the products.

From Fig.~\ref{fig:distribE} we generally find a very good agreement
between the MD results and the PST predictions. In the case of the
tetramer, the harmonic approximation for the vibrational DOS performs 
well. This is consistent with the existence of only a
single (stacked) isomer for this cluster at the energies considered
here. The Klots approximation for the rotational
density does not affect the results quantitatively. This indicates
that the results should be poorly sensitive to the exact shape of the
products. For the larger cluster (distributions not plotted) the agreement
remains very good but anharmonicities are found to be more significant.


The present agreement between molecular dynamics data and the PST
results is encouraging and allows us to fit the missing parameter
$C_0$ to reproduce the rate constant obtained from MD. Subsequently
the statistical properties can be calculated from PST for any
excitation energy. This is especially interesting at low energies, for
which molecular dynamics simulations are not practical. 

\subsubsection{Evaporation rates}
\label{sec:evtit}

\begin{figure*}[htbp]
\begin{center}
\begin{tabular}{cc}
\includegraphics[width=8.5 cm]{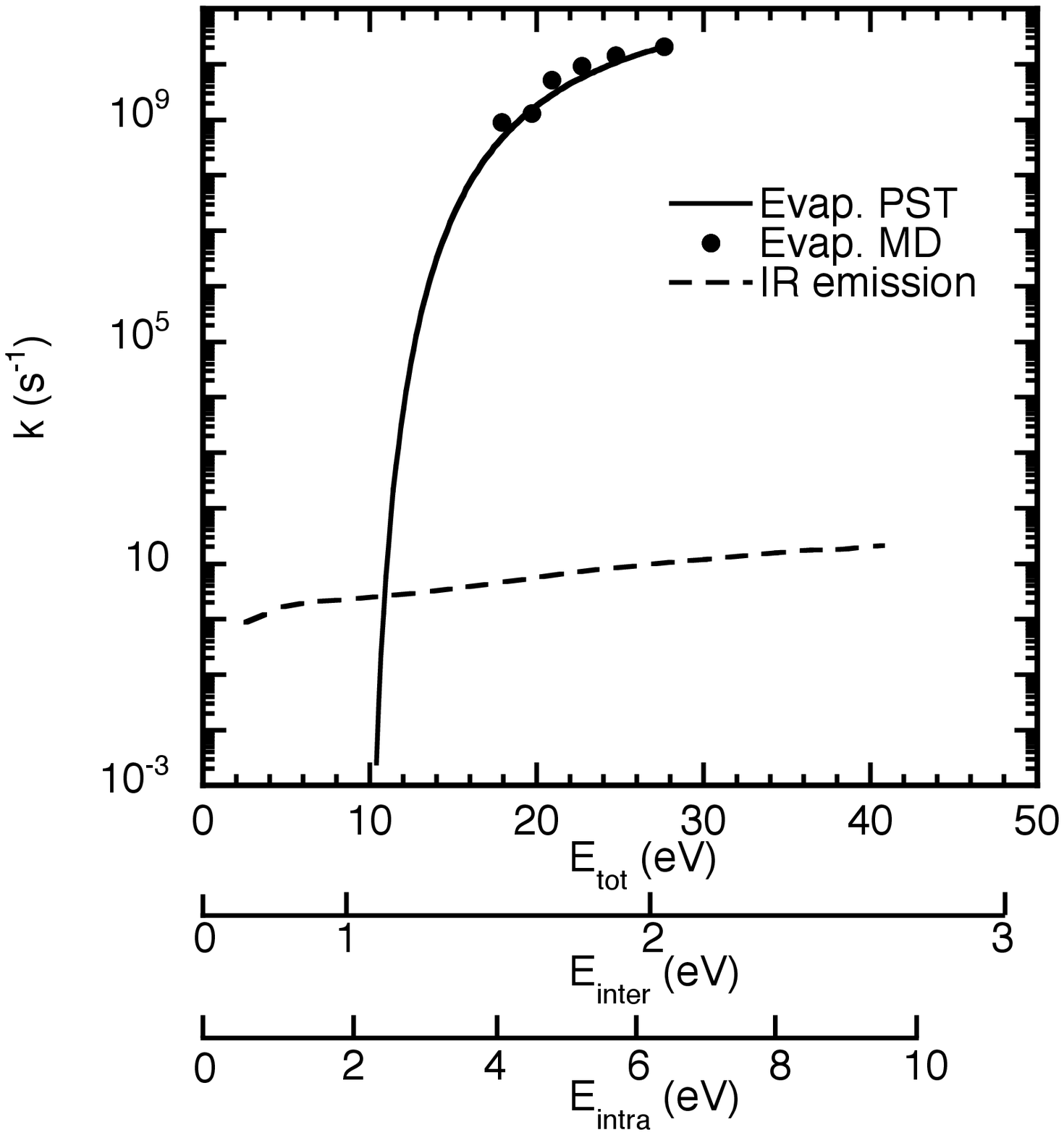}&
\includegraphics[width=8.5 cm]{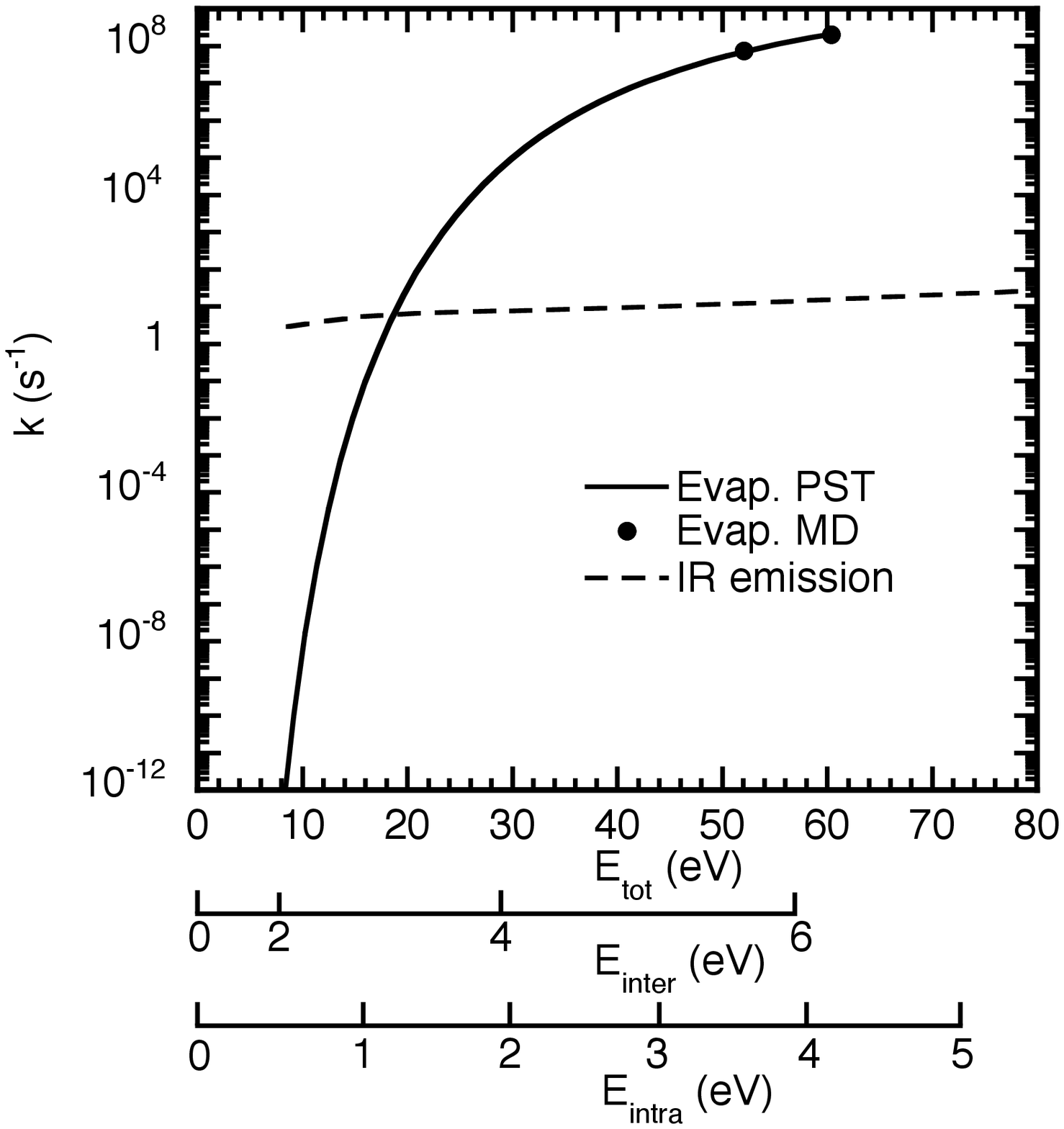}
\end{tabular}
\end{center}
\caption{Evaporation rates obtained with PST (solid lines) and MD
  simulations (symbols) as a function of the internal energy for
  coronene clusters with 4 (left panel) and 13 (right panel)
  molecules, respectively. The IR emission rates \citep{joblin02}
  are also represented as dashed lines. The total, intra- and intermolecular 
 energy scales  are related  
to each other through Eqn.~(\ref{eq:etotintra}) and (\ref{eq:etotinter}). }
\label{fig:kdeE}
\end{figure*}

The variations of the rate constant with internal energy are
represented in Fig.~\ref{fig:kdeE} for the two parent clusters.
The physical contents of these results will be discussed later in the
light of other competing decay channels.

\begin{figure}[htbp]
\begin{center}
\includegraphics[width=9 cm] {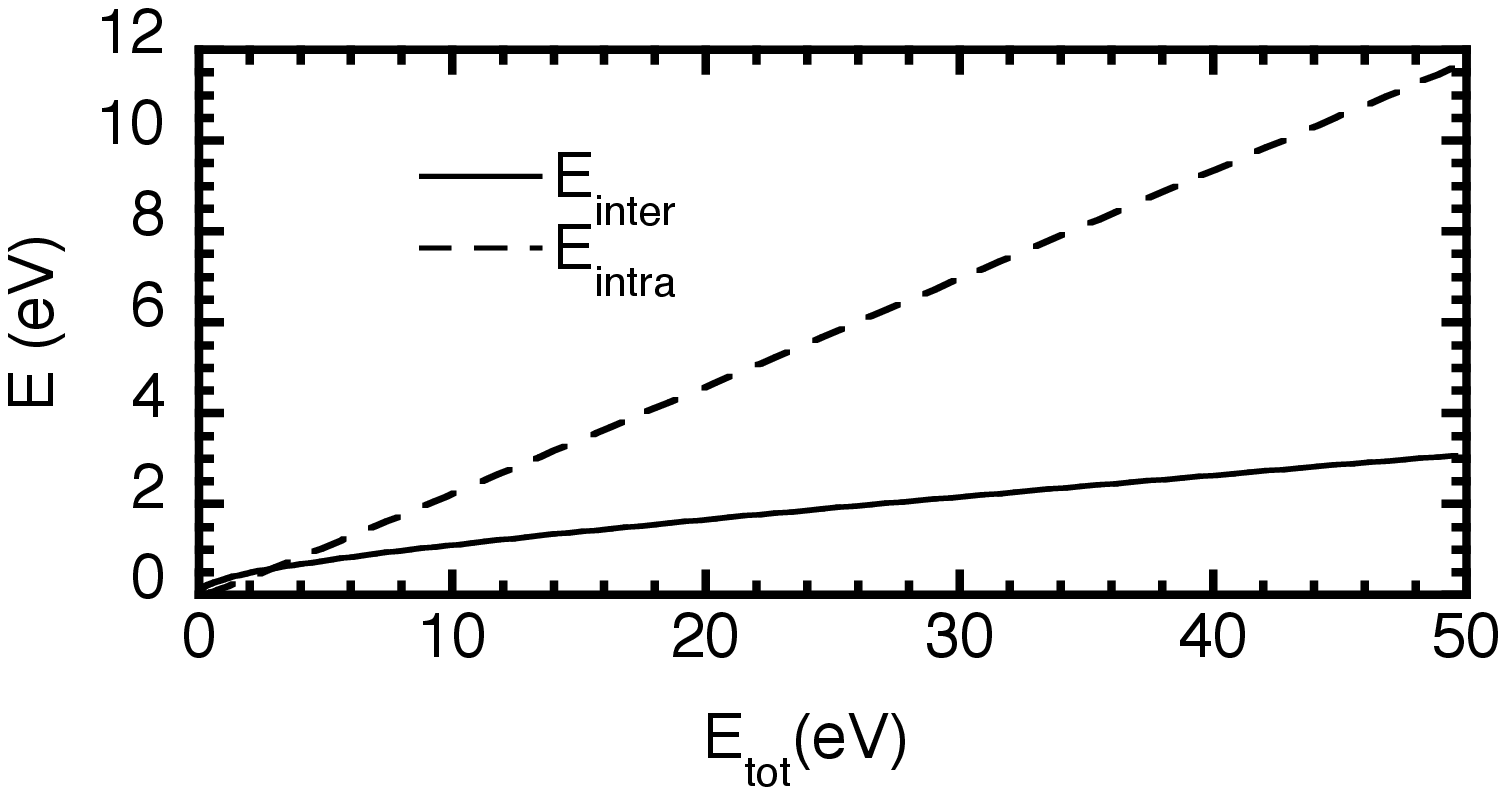}
\includegraphics[width=9 cm]{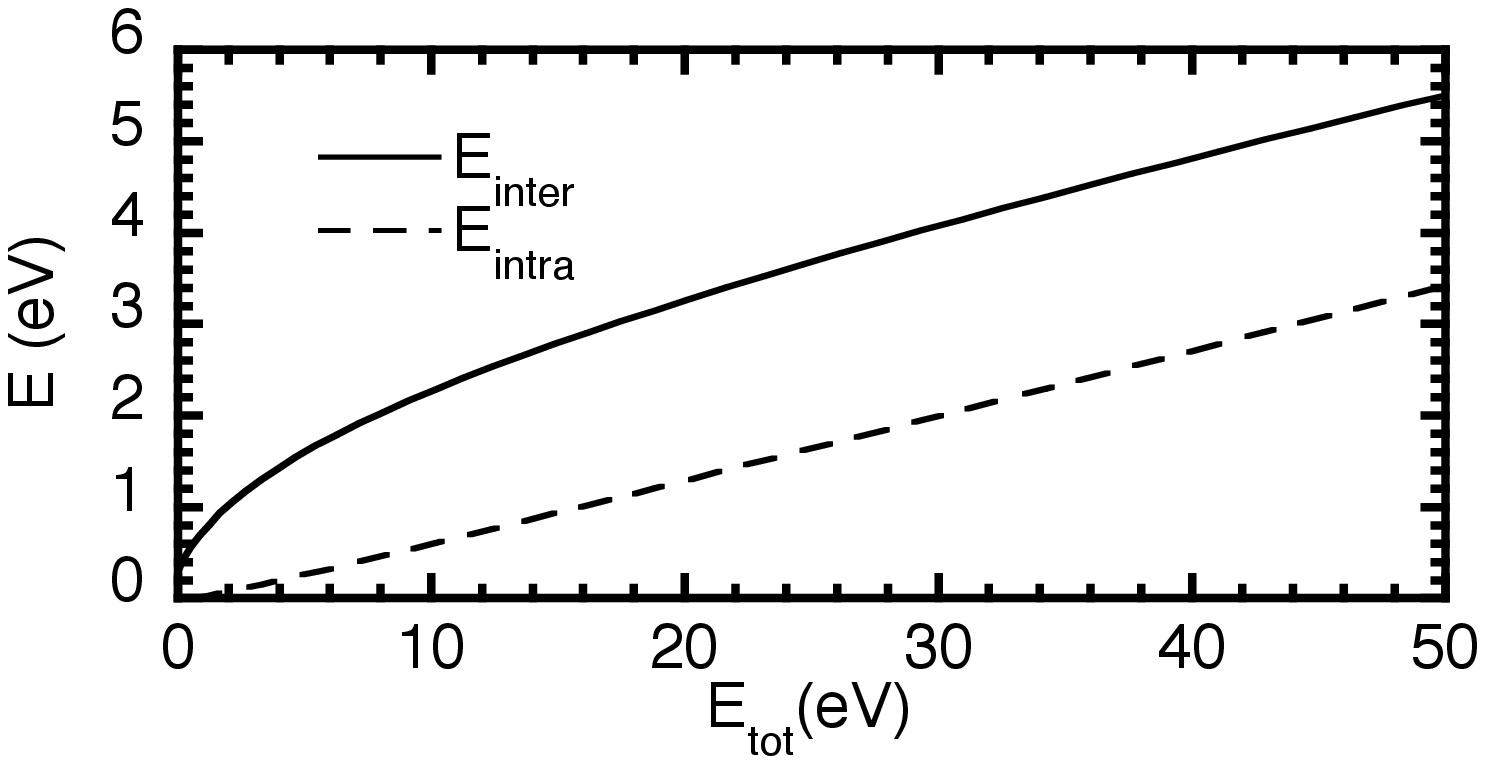}
\end{center} 
\caption{Intramolecular energy per molecule and total intermolecular energy
as a function of the total energy in the cluster for
(C$_{24}$H$_{12}$)$_{4}$ (top panel) and (C$_{24}$H$_{12}$)$_{13}$
(bottom panel), after removal of harmonic zero-point energies.}
\label{fig:rapener}
\end{figure}

An independent validation of phase space theory is obtained by comparing
 the rates from sets of molecular dynamics simulations
performed at different energies to the PST predictions. We thus performed 
additional
simulations for intermolecular energies in the 1.5--1.9~eV range for
(C$_{24}$H$_{12}$)$_4$ and at 5.5~eV for (C$_{24}$H$_{12}$)$_{13}$. 
The results shown in Fig.~\ref{fig:kdeE} are in very good
agreement with the statistical theory, even though a logarithmic scale
has been used for the graph. The discrepancy is found to be at most a
factor 1.85, which is satisfactory for a rate constant known to
span orders of magnitude.

\section{Competition between evaporation and IR cooling}
\label{sec:compet}

Infrared emission is a possible relaxation energy channel for an
excited cluster and can be in competition with evaporation. While
evaporation mainly involves the intermolecular energy, most of the IR
emission is driven by intramolecular modes. Therefore, in order to
compare the evaporation and IR emission rates, the inter- and
intramolecular energies must be first related to each other.

\subsection{Energy distribution}
\label{sec:edistrib}

In the next subsection, the spontaneous IR emission rates are
calculated for each molecule against their intramolecular energy
$E_{\rm intra}$. The evaporation rates as a function of intermolecular
energy $E_{\rm inter}$ are given in Sec. \ref{sec:evtit}. $E_{\rm
  intra}$ and $E_{\rm inter}$ both have to be expressed as a function of
the total cluster energy $E_{\rm tot}$. We first assume that the PAH
cluster remains at thermal equilibrium. Because most intramolecular modes
have a high frequency, the zero-point effects are important 
\citep{schmidt06}. 
The intermolecular modes, on the other hand, are all essentially soft and 
the intermolecular energy is equally distributed among those modes 
following classical equipartition. 

We calculate $E_{\rm intra}$ explicitly as a function of temperature
from the known vibrational frequencies $\{ \omega_i
\}$ of the single coronene molecule:
\begin{equation}
E_{\rm intra}(T)= \sum_i \hbar \omega_i \left(\frac{1}{2}+\frac{1}{\exp(\hbar
\omega_i/k_{\rm B}T) -1} \right),
\end{equation}
$k_{\rm B}$ being the Boltzmann constant.
The frequencies $\{ \omega_i \}$  are obtained from the DFT calculations of 
\citet{martin96b}  (which are more accurate than the present
TB model).

Similarly, we calculate $E_{\rm inter}$ as a function of temperature 
using the intermolecular frequencies from \cite{rapacioli05b}. 
The total energy of the cluster is the sum of  the intra- and
intermolecular energies. Figure \ref{fig:rapener} represents the
variations of the intramolecular energy per molecule $E_{\rm intra}$ 
and of the  total intermolecular energy  $E_{\rm inter}$
against $E_{\rm
tot}$. These curves were fitted using polynomial expressions of the form:
\begin{eqnarray}
E_{\rm tot} &=& \sum a_k (E_{\rm intra})^k \label{eq:etotintra} \\
&=& \sum b_k (E_{\rm inter})^k \label{eq:etotinter}
\end{eqnarray}
with fitting coefficients given in Table \ref{tab:fitpara}.

\begin{table}
  \begin{tabular}{|c|c|c|c|c|c|}
    \hline
Cluster & $a_0$ & $a_1$ & $a_2$ & $a_3$ & $a_4$ \\
\hline
(C$_{24}$H$_{12}$)$_4$ & 0.4345 & 4.2412 & 0 & 0 & 0 \\
(C$_{24}$H$_{12}$)$_{13}$ & 1.6614 & 13.9640 & 0 & 0 & 0 \\
\hline
Cluster & $b_0$ & $b_1$ & $b_2$ & $b_3$ & $b_4$ \\
(C$_{24}$H$_{12}$)$_4$ & -0.0338 & -1.0037 & 12.3750 & -3.2847 & 3.498 \\
(C$_{24}$H$_{12}$)$_{13}$ & -0.2084 & -0.2720 & 2.4862 & -0.1666 & 0.00449 \\
\hline 
\end{tabular}
  \caption{Parameters of the polynomial fits of  $E_{\rm tot}$,
  as a function of $E_{\rm intra}$ per molecule ($a_k$) and $E_{\rm inter}$ ($b_k$)
  for the (coronene)$_4$ and
  (coronene)$_{13}$ clusters.}
\label{tab:fitpara}
\end{table}

\subsection{Spontaneous IR emission rates and evaporation rates}
\label{sec:stability}

We address here the estimation of IR emission rates for neutral coronene clusters.
In a first approximation, the effects of clustering on the IR emission
of a specific molecule in the cluster will be neglected. Thus the
IR emission rate of the full cluster is obtained, considering the rates 
of individual molecules. 
The latter quantity explicitly depends on the energy
$E_{\rm intra}$. Calculations of the IR emission of coronene molecules in interstellar
conditions were reported by \cite{joblin02} [see also \citep{mulas06} for a more refined version
of this approach]. In this Monte Carlo code, the IR emission rates are described with microcanonical
statistics using an exact counting method for the density of states \citep{beyer}
based on harmonic oscillators.

To discuss qualitatively the competition between the evaporation and IR emission
processes,  a first convenient step is to define a mean emission rate.
This is achieved by summing the emission rates of all
IR active vibrational modes of all molecules in the cluster :
\begin{equation}
  k_{\rm IR}(E_{\rm intra})= \sum_n^{N_{\rm mol}} \sum_{i,j} k_{\rm
  IR}^{n,{\rm v_i}\rightarrow {\rm v_i}-1}(\nu_j)(E_{\rm intra}),
\end{equation}
where $k_{\rm IR}^{{\rm v}_i\rightarrow {\rm v}_i-1}(\nu_j)$ is the spontaneous
emission rate for a $\nu_j$ IR mode in the  ${\rm v}_i\rightarrow {\rm v}_i-1$
transition at the available energy $E_{\rm intra}$.

The variations of the mean IR emission rates $k_{\rm IR}$ and the
evaporation rates $k_{\rm evap}$ calculated in section \ref{sec:evtit}  for
(C$_{24}$H$_{12}$)$_4$ and (C$_{24}$H$_{12}$)$_{13}$ are represented
in Fig.~\ref{fig:kdeE} against $E_{\rm tot}$. The clusters are assumed
to be at thermal equilibrium, the energy repartition of section
\ref{sec:edistrib} was used.

A critical energy $E^*$ can be defined as the energy of equal
rates, $k_{\rm evap}(E^*)=k_{\rm IR}(E^*)$. This energy is
 10.4 eV for (C$_{24}$H$_{12}$)$_4$ and 18.6 eV for
(C$_{24}$H$_{12}$)$_{13}$.

While the evaporation rates increase very strongly (exponentially)
with $E_{\rm tot}$, the IR emission rates show much smaller,
polynomial-type variations. This suggests that the probability of
evaporating a cluster before IR cooling strongly decreases for
energies smaller than $E^*$ and becomes maximum for energies
larger than this threshold.

\section{Modelling the photophysics of a neutral coronene cluster in a radiation field}
\label{sec:photophys}
\subsection{Coronene cluster in a radiation field}

The Monte Carlo calculation
of \cite{joblin02} is based on the
exact stochastic method given by 
\cite{gillespie78} and
\cite{barker}. The method consists of propagating 'trajectories' that
are Markovian random walks over the energy levels of the
molecule. These trajectories depend on the following events:
absorption of UV photons, emission of IR photons, but also fragmentation,
as described in \cite{joblin06} . We have adapted this model to describe
the photophysics of coronene clusters in a specific radiation field.
The absorption rates were calculated using the
absorption cross section $\sigma$ of the coronene cluster and a
description of the astronomical radiation field. Since  $\sigma$ is
not known, we have scaled the cross section of the coronene molecule measured by
\cite{joblin92} by the number of molecules. The IR emission rates
as a function of $E_{\rm intra}$ were described in Sec. \ref{sec:stability}.
The evaporation rates were estimated by fitting the curves of
Fig. \ref{fig:kdeE} with $E_{\rm inter}$ calculated from E$_{\rm tot}$,
as explained in section \ref{sec:edistrib}.

The model allows us to calculate the time evolution of the internal energy
and the associated events (photon absorption, emission, or fragmentation).
Simulations are run over a given number of absorbed photons
(10$^6$ and 10$^7$ for the lowest and highest absorption rates, respectively).

\subsection{Simulation of NGC~7023. The  multiphoton absorption phenomenon}
\label{sec:ngc7023}

The photophysical evolution of (C$_{24}$H$_{12}$)$_{4}$ and
(C$_{24}$H$_{12}$)$_{13}$ clusters has been simulated in the
environment of the northern photodissociation region (PDR) of
NGC~7023. According to \cite{rapacioli05a},  different
regions corresponding to various depths inside the molecular
cloud were selected. The depth is quantified by the selective extinction $A_V$.

The radiation field in the different regions was calculated using the
Meudon  PDR code [\cite{lebourlot93}, \cite{lepetit06}]. The inputs for this calculation
 were modified with respect to previous calculations of
\cite{rapacioli05a}. For the total-to-selective extinction, we used a
value for $R_V$ of 5.5 according to the recent study by
\cite{witt06}. The stellar spectrum used for HD200775 is from the
Kurucz library with a temperature of 15{\,}000~K \citep{Kurucz91}.
The higher temperature derived by \citet{Ancker97} was found to be
inconsistent with the VUV flux observed on the star by the IUE satellite
[see \cite{valenti00} and IUE archive at http://www.vilspa.esa.es/iue/iue.html].
The radius of the star was taken as 10$R\sun$
which is typical for a Be star. We performed the same analysis as in
\cite{rapacioli05a} and inferred a hydrogen density of $7~
10^{3}$~cm$^{-3}$ and a value $A_V=0.41$ for the region of
maximum emission in the H$_2$ 9.7 $\mu$m band.

\begin{figure}[htbp]
\includegraphics[width=9 cm]{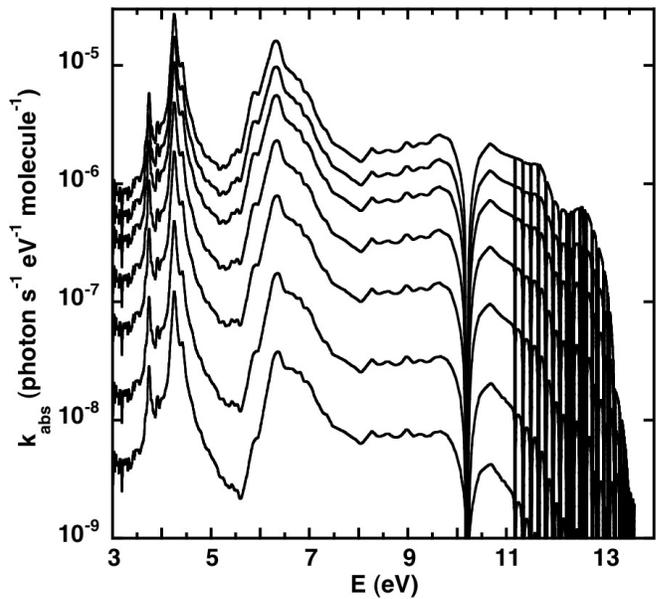}
\caption{Absorption rates of C$_{24}$H$_{12}$ at several depths inside
 the northern PDR of NGC~7023 : $A_V$= 0.41, 1.0,
1.6, 2.5, 3.5, 5.0 and 6.4 from top to bottom.
For clusters, the absorption rates
should be multiplied by the number of molecular units.}
\label{fig:absrate}
\end{figure}

The radiation field was calculated for $A_V$ values of 0.41, 1.0,
1.6, 2.5, 3.5, 5.0 and 6.4. The variations of the corresponding 
UV absorption rates are
represented in Fig. \ref{fig:absrate}. 

\begin{figure}[htbp]
\includegraphics[width=9 cm]{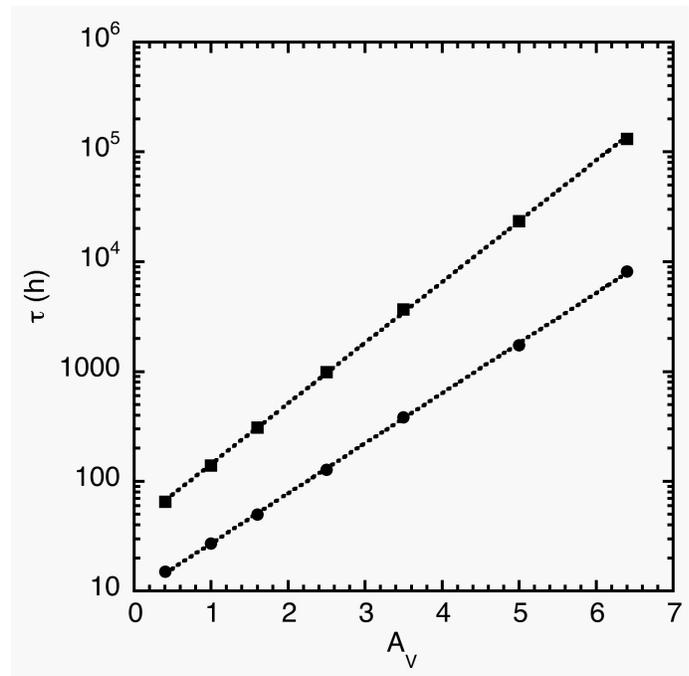}
\caption{Characteristic timescale for the evaporation of
  (C$_{24}$H$_{12}$)$_4$ (circles) and (C$_{24}$H$_{12}$)$_{13}$
  (squares) as a function of $A_{V}$ in the northern PDR of
  NGC~7023.}
\label{fig:evaptime}
\end{figure}

\begin{figure}[htbp]
\includegraphics[width=8.5 cm] {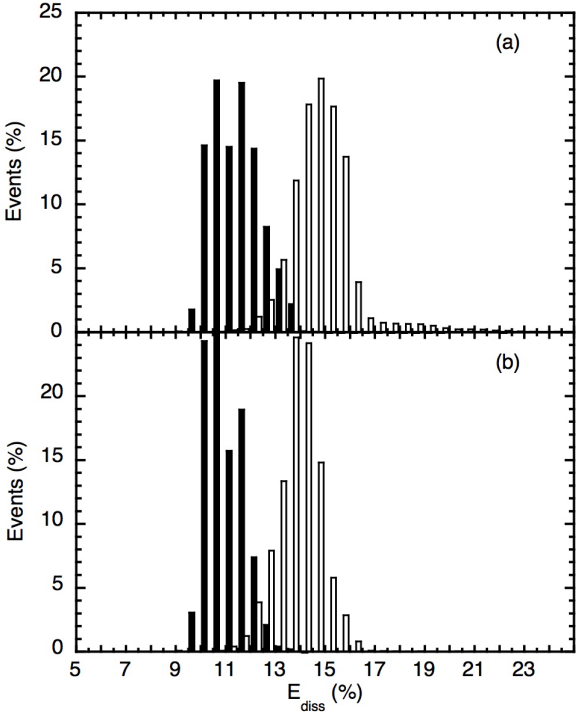}
\caption{Normalized distributions of the internal energy in (C$_{24}$H$_{12}$)$_4$
and (C$_{24}$H$_{12}$)$_{13}$ clusters prior to their evaporation
(black and white bars respectively).
The calculations were performed in two different 
radiation fields at
(a) $A_V=0.41$ and (b) $A_V=5.0$.}
\label{fig:eavtevap}
\end{figure}

\begin{figure}[htbp]
\includegraphics[width=9 cm] {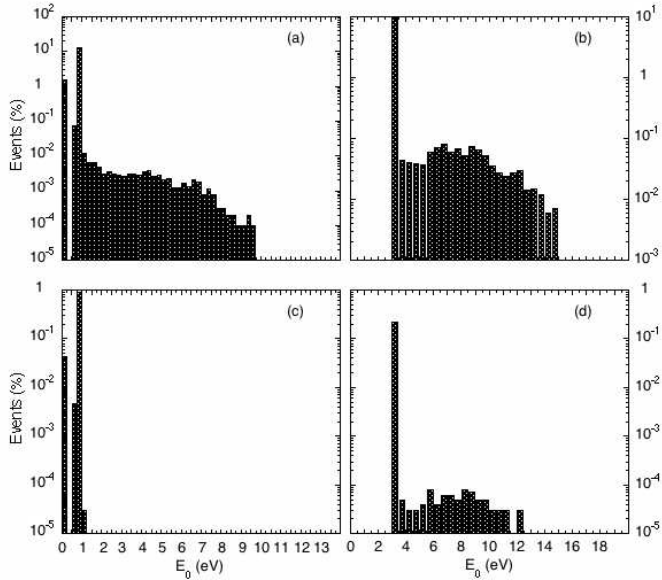}
\caption{Internal energy distributions of the clusters before absorption of the VUV
photon leading to dissociation of (C$_{24}$H$_{12}$)$_4$ (left)
and (C$_{24}$H$_{12}$)$_{13}$ (right). The calculations were
performed in two different radiation fields at
(a and b) $A_V=0.41$ and (c and d) $A_V=5.0$. The distributions are normalized to
the total number of absorbed UV photons.}
\label{fig:eavtabsevap}
\end{figure}

The average timescales of evaporation are shown in Fig. \ref{fig:evaptime}
for (C$_{24}$H$_{12}$)$_4$ and (C$_{24}$H$_{12}$)$_{13}$.
While the small evaporation timescale for  (C$_{24}$H$_{12}$)$_4$ can be
explained by the low critical energy $E^*$ of this cluster
(see section \ref{sec:stability}), the evaporation of the larger (C$_{24}$H$_{12}$)$_{13}$ has
its critical energy above the cut-off of the radiation field.
This is illustrated in Fig. \ref{fig:eavtevap}, which displays the distributions of
internal energy of the clusters prior to their evaporation.
The mean energy for dissociation is in the range 10.5--11.5~eV for (C$_{24}$H$_{12}$)$_4$
 and $\sim$14.5~eV for (C$_{24}$H$_{12}$)$_{13}$.
This value is slightly shifted towards lower energies at $A_V=5.0$ because of the
extinction, which preferentially affects the more energetic photons.
Another effect that can be observed in Fig.\,\ref{fig:eavtevap} is the absence of the
high-energy tail for $A_V=5.0$. 
Such events are only 
possible if the molecule can absorb two VUV photons without 
totally cooling by IR emission  between the two absorptions, which requires in turn requires 
sufficiently high absorption rates.

From Fig.~\ref{fig:evaptime}, it is clear that (C$_{24}$H$_{12}$)$_{13}$ clusters have accumulated the energy of more
than one photon to reach the dissociation threshold, even in the more
embedded regions. Figure\,\ref{fig:eavtabsevap} shows the internal energy contained in the
clusters before absorption of the VUV photon leading to dissociation. In the case of the
large clusters (13 units), a clear threshold is seen at 3~eV.
This threshold is very sharp at low energies, as could be anticipated: there is 
no event in the 2.85~eV channel
of the histogram and 60--70\% of the total number of events are found in the 3.0~eV channel. This fraction
drops very quickly at higher energies with $\sim$30\% in the 3.25~eV channel and the remaining
1--10\%
spread over channels up to 15.5~eV and 13.5~eV for $A_V$ of 0.41 and 5.0, respectively.

The peak in the 3.0~eV channel can
be interpreted as the consequence of truncating the mean energy
stored in the clusters at the threshold where the clusters are in equilibrium with the radiation field. 
Indeed, the model shows that a full dissipation of the energy absorbed by IR emission cannot be
achieved before another photon is absorbed.
At low internal energies, the cooling efficiency decreases as only low frequency modes with
weak intensities are involved.

Contrary to large clusters, (C$_{24}$H$_{12}$)$_4$ species can be
evaporated by a single VUV photon, multi-photon events playing only a minor
role, as seen in Fig.\,\ref{fig:eavtabsevap}.
In this case, the 0~eV channel corresponds to clusters that evaporate
as soon as they absorb their first UV photon. 
The peak around 0.8\,eV corresponds to the mean
energy stored in equilibrium with the radiation field.
In this case, we can conclude from the observation of a population in the 0~eV energy channel
that accumulation of energy is not required for the evaporation of (C$_{24}$H$_{12}$)$_4$ clusters.
This is obvious since energies lower than 13.6~eV are already required.

\section{On the competition between formation and destruction of PAH
  clusters under interstellar conditions}
\label{sec:balance}
\subsection{Growth constants}

We estimate here the rate constant of the nucleation process
associated with the growth of a single molecule on a  cluster of
$i-1$ molecules, both species being neutral.
In the interstellar medium, the molecular gas can be considered as
very rarefied, hence molecules and clusters have enough time to release
their internal energy via IR emission before any collision occurs.
In addition, we will assume that the molecular gas is thermalized, 
Boltzmann laws governing the distribution of translational and
rotational energies of both clusters and single molecules.

If $\delta \tilde{n}_i$ represents the number of clusters of size $i$
formed upon the nucleation reaction $1+ (i-1)\rightarrow i$ per unit of
time $\delta t$ and inside a unit of volume, the growth constant
$k_i$ is defined such that: 
\begin{equation}
\frac{\delta \tilde{n}_i}{\delta t}=k_i n_1 n_{i-1}.
\end{equation}
In this equation $n_1$ and $n_{i-1}$ are the densities of single
molecules and clusters of size $i-1$ per volume unit, respectively. 
$k_i$ can be expressed from standard collision theory \citep{lifshitz} as: 
\begin{eqnarray}
k_i&=&\int\!\!\!\int\!\!\!\int f_1({\rm v}_1)f_{i-1}({\rm v}_{i-1}) \|{\bf v}_r\| 
\sigma_0 \nonumber \\ &&\displaystyle \times 
p_{i,+}({\rm v}_r;E_{\rm r}) d^3{\bf v_1} d^3{\bf v_{i-1}} f(E_{\rm r}) dE_{\rm r}.
\label{eq:ki2}
\end{eqnarray}
In the above equation, $f_1({\rm v}_1)$ and $f_{i-1}({\rm v}_{i-1})$ are the probability
densities of the incident molecule and the cluster, which depend on
their respective velocities ${\rm v}_1$ and ${\rm v}_{i-1}$ only. $f(E_{\rm r})$ is the
probability distribution  of the rotational energy available at the
beginning of the collision. $\sigma_0$ is the nucleation cross
section, taken here as the geometrical cross
section. $\bf{v}_r=\bf{v}_1-\bf{v}_{i-1}$ is the relative
velocity vector. $p_{i,+}({\rm v}_r;E_{\rm r})$ is the probability that the collision
leads to the formation of cluster size $i$. Assuming that $f_1({\rm v}_1)$ and
$f_{i-1}({\rm v}_{i-1})$ both follow a Boltzmann law at temperature $T$, Eq.~(\ref{eq:ki2}) can be
written more explicitly as:
\begin{eqnarray}
k_i&=&\displaystyle \left(\frac{\sqrt{m_1m_{i-1}}}{2\pi k_{\rm B}
  T}\right)^3 \int\!\!\! \int \|{\bf v}_r\|p_{i,+}({\rm v}_r;E_{\rm r}) \sigma_0
  \nonumber \\ &&\displaystyle \times
  \exp\left[-\frac{(m_1+m_{i-1}){\rm v}_G^2+\mu {\rm v}_r^2}{2k_{\rm B} T}\right]
  d^3{\bf v}_G d^3{\bf v}_r.
\label{eq:ki3}
\end{eqnarray}
In the latter equation, $m_1$ and $m_{i-1}$ are the masses of the incident
molecule and the cluster, respectively; ${\rm v}_G$ is the velocity of the
centre of mass of the entire system. 
The integration of this equation over ${\rm v}_G$ gives: 
\begin{equation}
k_i(T)= \frac{\sqrt{8/\pi\mu}}{(k_{\rm B} T)^{3/2}} \int\!\!\! \int  
E_{\rm t} \sigma_0 p_{i,+}(E_{\rm t};E_{\rm r}) e^{-E_{\rm t}/k_{\rm B} T}dE_{\rm t} f(E_{\rm r}) dE_{\rm r},
\label{eq:ki4}
\end{equation}
where $E_{\rm t}$ is the translational energy in the centre of mass reference frame,
and $\mu$ the reduced mass of the system.

In order to calculate quantitative values of $k_i$, the collision
cross section is approximated as the geometric area of the PAH
molecule (7.1~10$^{-19}$ $\AA^{-2}$).
In section \ref{sec:resag} 
we only calculated values of $p_{i,+}(E_{\rm t};E_{\rm r})$
for the specific case $E_{\rm r}=E_{\rm t}$.
However a lower bound to $k_i$ can be estimated as follows.

At a given translational energy $E_{\rm t}$, the cluster growth probability 
decreases with the increase of rotational energy $E_{\rm r}$, as it implies
an increase in the total energy. Hence
\begin{equation}
  p_{i,+}(E_{\rm t};E_{\rm r}\geq E_{\rm r}^0)\leq p_{i,+}(E_{\rm
  t};E_{\rm r}^0). 
\end{equation}
Similarly, at a given  $E_{\rm r}$, the cluster growth probability 
decreases with increasing $E_{\rm t}$:
\begin{equation}
  p_{i,+}(E_{\rm t}\geq E_{\rm t}^0;E_{\rm r})\leq p_{i,+}(E_{\rm
  t}^0;E_{\rm r}).
\end{equation}
These two trends have been checked using explicit atomic simulations,
and performing a few tens of molecular dynamics trajectories of the
nucleation process. A lower bound to $k_i$ can be calculated assuming
that
\begin{equation}
  p_{i,+}(E_{\rm t};E_{\rm r})= \cases{ p_{i,+}(E_{\rm t};E_{\rm t}) & if $E_{\rm t}\geq
    E_{\rm r}$; \cr
p_{i,+}(E_{\rm r};E_{\rm r}) & if $E_{\rm t} < E_{\rm r}$.}
\end{equation}
Moreover, the collision constants $p_{i,+}$ provide an upper bound to
the growth constants, by taking $p_{i,+}=1$ in Eq.~(\ref{eq:ki4}).

\begin{figure}[htbp]
\includegraphics[width=9 cm] {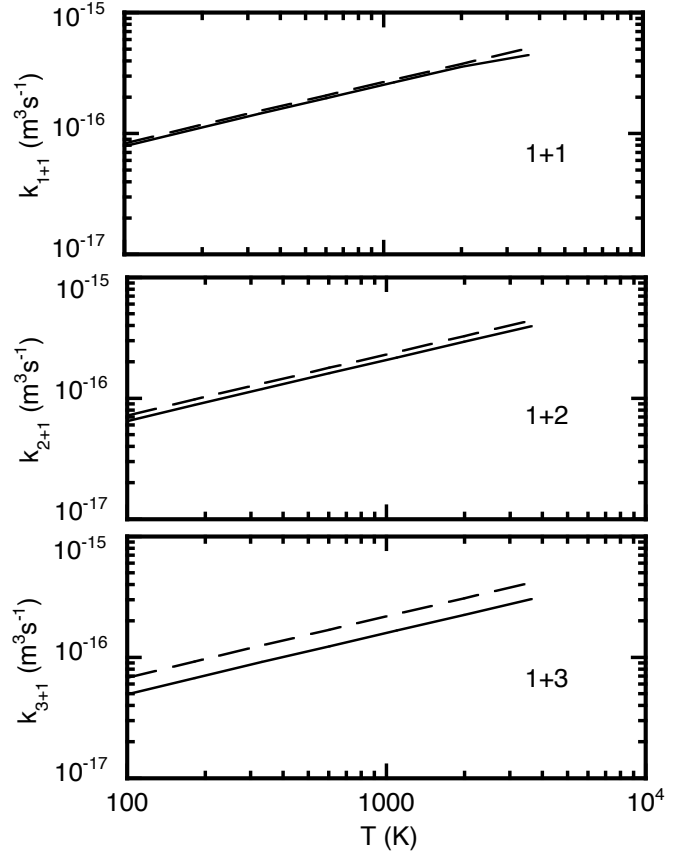}
\caption{Collision rates (dashed lines) and growth rates (solid lines)
for the nucleation processes $1 + 1 \rightarrow 2$, 
$1+2 \rightarrow 3$ and $1+3 \rightarrow 4$ plotted as a function of
temperature. }
\label{fig:kcol}
\end{figure}

Figure~\ref{fig:kcol} shows the lower and upper limits to $k_i$ with
$i=2,3,4$ obtained in the temperature range 10$^2$--10$^4$~K.
The two limits appear to be very close. This is due to the fact that, except
for extremely high collisional energies, nearly all collisions 
lead to effective growth. In a first approximation, the growth rate 
can then be taken as the collision rate. We also note that the growth
constant increases monotonically with temperature. Indeed a
temperature increase enhances the rate of collisions which, for
$T<2000$~K, nearly always lead to nucleation.
Similar calculations have also been performed with the probability $p_{i,-}$
of destroying the parent cluster. The corresponding rates are found to
be a few orders of magnitude smaller than the growth rates, hence we
conclude that this process can be neglected.

\subsection{Competition between formation and destruction in
    neutral PAH clusters}
\label{sec:discussion}
In their study of the reflection nebula NGC 7023,
\cite{rapacioli05a} gave hints that PAH molecules are likely to be 
condensed as clusters.
These clusters are evaporated at the border of the cloud
where isolated PAHs are observed. A minimum size of 400 carbon atoms
for the observed clusters was also inferred from 
the observations.

An order of magnitude for the typical timescale of PAH clustering in
this reflection nebula can be estimated from the calculations
reported in section \ref{sec:balance}. We use a density $n_{\rm
H}=7~10^3$~cm$^{-3}$ for NGC 7023, as discussed previously 
(sec. \ref{sec:ngc7023}),
and a carbon abundance $n_{\rm C}=2~10^{-4}n_{\rm H}$
following \cite{sofia04}. We consider that
about 20\% of the carbon is contained in PAH molecules.
For simplicity, and in order to apply our previous results, we also
assume that all PAH molecules are coronene. This leads to a density of
coronene molecules of $1.12~10^{-2}$~cm$^{-3}$.
According to the PDR model, the kinetic temperature varies from 500~K
to 60~K in the range $A_V=0.41$--6.4.

We find that the aggregation timescale is typically $10^4$--$8~10^4$ years.
This is much longer than the photoevaporation timescales derived for
(C$_{24}$H$_{12}$)$_4$ and (C$_{24}$H$_{12}$)$_{13}$ in section
\ref{sec:ngc7023}, which are below  $1.5~10^5$ hours (17 years).
Therefore, these clusters should be destroyed by the
UV flux much faster than they can be reformed. This also implies that, when a
large cluster starts to evaporate, the small fragments are also likely
to photoevaporate themselves. It can be deduced from this theoretical work that, in the case
of NGC~7023, the minimum size of the coronene clusters should be larger
than 13 molecular units  (312 carbon atoms), which is
consistent with the minimum size of 400 carbon atoms inferred from the
observations \citep{rapacioli05a}.


\subsection{Effects of ionization}
\label{sec:iondensite}

Until now we have focused on the description of neutral clusters. Our
model can be used to constrain the size distribution of PAH clusters
as a function of depth in the molecular cloud. One would like to be able to constrain the charge as well, similarly to what is done for single PAHs
\citep{Bakes01,LiDraine02}. While we can give hints about charge 
effects, fundamental studies are clearly required before considering
further modeling.

Even though PAHs are likely to carry a charge in many objects of the
ISM \citep{Hudgins99,Allamandola89}, \citet{rapacioli05a} 
have shown that the different populations of very small dust particles
are spatially decorrelated in two PDRs
 NGC~7023 and $\rho$Oph-SR3. As the
distance from the stars increases, the mid-IR emission is found to be
dominated first by positively ionized PAHs, then by neutral PAHs, and
finally by VSGs. Similarly to single PAHs, PAH clusters could carry a
charge. We expect this charge to be most likely positive, since the
lifetime of anions will be limited by the easy photodetachment in
these UV-irradiated regions where PAH clusters can be observed by
their mid-IR emission. Charge effects can affect both the formation
rate and the evaporation rate for different but related reasons.

Let us first consider the case of multiply charged systems. The
nucleation between two charged subsystems seems very unlikely due to
the strong Coulomb repulsion. 
We thus exclude the reactions of positively charged clusters with PAH
cations. The stability of an existing assembly of PAH molecules
will critically depend on its charge state, in addition to its
size. Other factors, including the size of the PAH itself and the
excitation mechanisms, presumably play an important role as well. 
In a multiply ionized
cluster, Coulomb fragmentation will generally enhance the possibility
of emitting single (charged) molecules. In addition, new fragmentation
channels will be open if a single PAH molecule carries more than
one charge, and various intermolecular and/or intramolecular
Coulomb multifragmentation processes could take place. While the
photofragmentation dynamics of a PAH assembly (or even a single PAH
cluster) would be worth investigating in its own, it is
beyond the scope of the present paper.

 Based on previous theoretical 
calculations on small singly charged clusters
of aromatic molecules
such as benzene, naphtalene or anthracene \citep{piuzzi,Bouvier02},
cationic clusters are expected to be significantly more stable than
their neutral counterparts from the energetic point of view. This
increased stability reflects the charge delocalization over several
molecules in the cluster, a collective effect which can{\em not}\/ be
described using simple explicit potentials. Because the evaporation 
rate constant varies approximately exponentially
with the dissociation energy, cationic clusters will be significantly
more resistant to the decay of a neutral molecule.

In the region where PAH clusters are observed, PAHs are expected to 
be mainly neutral \citep{rapacioli05a}, and we can limit our discussion
to the case where a neutral molecule 
collides with a (PAH)$_n^+$ 
cluster. For the nucleation rate, the PAH molecule being nonpolar, the
long-range ion/neutral interaction scales as $-C_4/r^4$ where the 
constant $C_4$ is proportional to the electronic dipole polarizability
$\alpha$ of the neutral PAH. In a very simple approach, we use a
Langevin model to estimate the collision rate constant. Based on our
previous results on neutral/neutral collisions, we can reasonably assume that
all collisions between a neutral PAH and a cationic cluster lead to an
effective growth of the cluster. The Langevin prediction for the
nucleation rate is then given by $k_{\rm nucl}=2 \pi e q
(\alpha/\mu)^{1/2}$ [see for instance \cite{Herbst02}], 
where $e$ is the electronic charge, q the ionization degree 
($q=1$ for singly-charged clusters)
 and $\mu$ the reduced mass of the reactants. An average
polarizability of the coronene molecule, as estimated from the 
components of the polarizability tensor via density-functional theory
calculations \citep{malloci_private}, is $\alpha=46~10^{-24}$~cm$^3$.
Using this value the nucleation rate is approximately $k_{\rm
nucl}=9~10^{-10}/q$~cm$^3$s$^{-1}$. 
Together with the PAH density
derived previously for NGC~7023, this leads to aggregation timescales of
[2.0--3.5]/$q~10^3$ years.

For the specific case of NGC~7023, considered in section
\ref{sec:discussion}, the conclusion that small clusters are destroyed
faster than they can be reformed could thus be modified if the
clusters carry one single positive charge, as this would decrease
their evaporation rate (mostly through increasing their binding
energy) but also enhance the nucleation rates.

\subsection{Astrophysical implications}
\label{sec:astro}

As seen above, a single positive ionization is expected to stabilize PAH
clusters with respect to evaporation, and to favor their formation by
nucleation of a neutral molecule. Our conclusion that neutral PAH clusters
are evaporated efficiently in the northern PDR of NGC~7023 without
being reformed could thus be revised in the case of cationic clusters.

The linear scaling of the growth constant with inverse density
suggests another mechanism that could affect the conditions of reformation.
The northern PDR of NGC~7023 is heterogeneous, with some
high-density filaments ($n\sim 10^5$--$10^6$~cm$^{-3}$) embedded in a
more diffuse medium with $n\sim 10^4$~cm$^{-3}$ [\citet{fuente2000}
and references therein]. With a value of $5\,10^5$~cm$^{-3}$ for the
density, the timescale for formation could be lowered to  140--700
years for a neutral cluster, and to 30--50 years for an ionized
cluster. Considering the dissociation rates reported in
Fig.~\ref{fig:evaptime} for neutral species, and taking into account
the extra stability of the cationic species, the PAH clusters could 
thus survive much longer and closer to the border of the
cloud. A scenario in which the growth of PAH clusters occurs in the PDR
itself is however unlikely because it 
would involve the survival of dimers in PAH-emitting regions. 
This suggests that such clusters would be formed
in the UV-shielded parts of the cloud. There is time to form them
during the cycle of matter from diffuse to dense clouds. Indeed the 
quiet phase of molecular clouds will typically last 10$^6$ years, 
allowing clusters of tens of molecules to be formed. 
The formation of young stars in these clouds creates
PDRs, which are fed in with clusters and rapidly submitted to
photoevaporation.
Dynamical effects in PDRs such as an advancing
photodissociation region \citep{Lemaire99,fuente99,fuente2000} could
then bring these clusters into the UV-irradiated regions.

\section{Conclusion}
\label{sec:con}

The formation and physico-chemical evolution of PAHs is a question of
great interest in astrochemistry. \cite{rapacioli05a}
presented strong evidence that  PAH clusters are good candidates for 
small carbonaceous grains. 
These authors suggested that those grains are indeed PAH clusters, and
a minimum size of 400 carbon atoms per cluster was inferred in
NGC~7023. In this paper, the competition between the formation and
destruction processes of such PAH clusters in the neutral state was
investigated theoretically under realistic astrophysical conditions.
Coronene clusters have been considered here as representative
prototypes of PAH clusters.

The growth of neutral PAH clusters by nucleation of individual molecules
was simulated using a model combining an explicit atomic force
field to describe the intermolecular interactions, along with a quantum
tight-binding approach for the intramolecular interactions. It was
found that, under interstellar conditions, most of the collisions lead
to cluster growth. This was interpreted as due to the large number of
intramolecular modes that contribute to absorbing the excess
collision energy.

Evaporation was treated in the framework of phase space theory,
supplemented with molecular dynamics simulations. The clusters
(C$_{24}$H$_{12}$)$_{4}$ and (C$_{24}$H$_{12}$)$_{13}$ were
investigated, and a good agreement between the simulation results and
the predictions of the statistical theory was found for the
distribution of kinetic energy released after evaporation.
The calculated evaporation rates were then introduced in
an infrared emission model based on microcanonical statistics. This
model allowed the simulation of these clusters in a specified interstellar
radiation field.
 
When applied to the reflection nebula NGC~7023, our model indicates that
(C$_{24}$H$_{12}$)$_{4}$ and (C$_{24}$H$_{12}$)$_{13}$ are 
both photoevaporated much faster than they can be reformed. 
In the case of (C$_{24}$H$_{12}$)$_{13}$, this evaporation is possible
if the UV absorption rate is faster than the IR emission rate, so that
at least $\sim$3~eV of internal energy are stored in the cluster before
absorption of a VUV photon. These results provide a plausible
explanation for the absence of small VSGs observed in this PDR. They
also suggest that PAH clusters are formed inside the molecular cloud
and transported in the PDR by dynamical effects.

The present investigation is a first step towards the
treatment of the complex dynamics of nucleation, thermal and
photo fragmentations of PAH clusters in the interstellar medium.
Our work could be extended by limiting the approximations done
in the calculations of aggregation rates (impact parameter and initial
energy distribution) and of evaporation rates (mainly the angular
momentum). A next step could be to study larger clusters, especially
those beyond the 400 carbon atom limit. It would be
important to determine whether such large clusters could survive
inside a reflection nebula like NGC~7023. Unfortunately, the
computational efforts for simulating the evaporation of large clusters
by molecular dynamics, to calculate their vibrational densities of
states with Monte Carlo simulations, or to apply the IR/evaporation model will
all tend to become extremely costly, at least one order of magnitude
higher than in the present study. Further approximations will then become
unavoidable for a proper extension of the present work.

Two important factors, namely ionization and density
heterogeneities, might enhance the stability of PAH clusters in
PDRs. These factors have been qualitatively discussed here.
Only singly-charged cationic clusters are expected to be more stable
than neutrals. Conversely, 
multicharged species and anionic clusters should be less stable due to
electrostatic repulsion and electronic photodetachment processes,
respectively. For the formation problem,
it would be very useful to build a complete model to
describe singly charged cationic assemblies of PAH molecules.

To be chemically relevant, such a model would need to include
two main effects, that are not important for neutral clusters.
Firstly, the charge will lead to $1/r^4$ polarization contributions,
or even many-body implicit interactions between induced dipoles at
higher order. Secondly, the model should account for charge
delocalization, the missing electron being able to hop between
neighbouring molecules. In homogeneous molecular systems, the charge
is not restricted to be carried by a single PAH, but may extend over
several units, typically between two and four \citep{piuzzi}. This
phenomenon is known as charge resonance. Obviously, polarization and
charge delocalization effects are not unrelated, but their many-body
and quantum characters make them significantly harder to describe than
the pairwise repulsion-dispersion-electrostatic interactions in the
presently investigated neutral clusters.

Thus, for cationic PAH clusters a more complex atomic model is
necessary to take into account electron tranfer. A model along those
lines was developed by \cite{piuzzi}  and \cite{Bouvier02} in order to investigate
the structure of small clusters of aromatic molecules. The extension
of this approach to large clusters and/or large PAH molecules will
be computationally demanding, considering the extensive sampling
required in the study of unimolecular evaporation. In the context
of nucleation, the coupling between intermolecular and intramolecular
modes will further burden the application of this model.
 With respect to neutral clusters, the theoretical modeling of 
cationic PAH clusters is therefore expected to be more involved. 
Efforts in this direction will be invaluable to improve our understanding
of the physical and chemical evolution of photodissociation regions.

\begin{acknowledgements}
 We thank J. Le Bourlot for help with the PDR code
and M. Schmidt for useful discussions. We gratefully acknowledge
the anonymous referee for constructive comments. This work was
supported by the GDR 2758 ``Agr\'egation, fragmentation et
thermodynamique des syst\`{e}mes complexes isol\'{e}s.'' 
We greatly thank the financial support from the Alexander von Humboldt Foundation. 
We also thank IDRIS for a generous allocation of computater resources.
\end{acknowledgements}

\bibliographystyle{aa}
\bibliography{rapacioli_astroph.bib}

\end{document}